\documentclass[aps,reprint,superscriptaddress,nofootinbib,longbibliography]{revtex4-2}

\usepackage[utf8]{inputenc}
\usepackage{amsmath}
\usepackage{amsthm}
\usepackage{amsfonts}
\usepackage{amssymb}
\usepackage{mathrsfs}
\usepackage{graphicx}
\usepackage{color}
\usepackage{enumerate}
\usepackage{tabularx}
\usepackage{tensor}
\usepackage{bm}
\usepackage{subcaption}
\usepackage{tikz}
\usepackage{float}

\begin{document}
	
\title{Stable Cosmology from Minimal Theory of Mass-Varying Massive Gravity}
	
\author{Ahmad Khoirul Falah}
\email{akfalah49@gmail.com}
\affiliation{Theoretical Physics Laboratory, Theoretical High Energy Physics Research Division, Faculty of Mathematics and Natural Sciences, Institut Teknologi Bandung, Jl.~Ganesha no.~10 Bandung, Indonesia, 40132}
	
\author{Andy Octavian Latief}
\email{latief@itb.ac.id}
\affiliation{Physics of Magnetism and Photonics Research Division, Faculty of Mathematics and Natural Sciences, Institut Teknologi Bandung, Jl.~Ganesha no.~10 Bandung, Indonesia, 40132}
	
\author{Husin Alatas}
\email{alatas@apps.ipb.ac.id}
%\affiliation{Indonesian Center for Theoretical and Mathematical Physics (ICTMP)}
\affiliation{Theoretical Physics Division, Department of Physics, IPB University (Bogor Agricultural University), Jl.~Meranti, Kampus IPB Darmaga, Bogor 16680, Indonesia}
	
\author{Bobby Eka Gunara}
\email{bobby@itb.ac.id (Corresponding author)}
\affiliation{Theoretical Physics Laboratory, Theoretical High Energy Physics Research Division, Faculty of Mathematics and Natural Sciences, Institut Teknologi Bandung, Jl.~Ganesha no.~10 Bandung, Indonesia, 40132}
%\affiliation{Indonesian Center for Theoretical and Mathematical Physics (ICTMP)}
	
\begin{abstract}
	We study cosmological perturbations in the minimal theory of mass-varying massive gravity (MTMVMG), a constrained extension of mass-varying massive gravity that propagates only three physical degrees of freedom. We show that MTMVMG admits a stable cosmological solutions i.e. free from ghost, gradient, and tachyonic instabilities around the homogeneous and isotropic background. We further demonstrate that the dynamical external scalar field$\textendash\textendash$which is responsible for the mass of the graviton$\textendash\textendash$can suitably serve as either dark energy or the inflaton, yielding a description consistent with current cosmological observations.
\end{abstract}

\maketitle
	
%
%A central requirement for any viable cosmological model is the stability of its background evolution. In cosmological perturbation theory, this involves analyzing the behavior of small fluctuations around a given background spacetime, which is often assumed to be homogeneous and isotropic. These perturbations are not only critical for ensuring theoretical consistency but also underpin the formation and growth of cosmic structures, such as galaxies and clusters. While general relativity (GR) provides an excellent description of gravity on astrophysical scales, its extrapolation to cosmological distances faces tension with several observations$\textendash\textendash$most notably the accelerated expansion of the Universe. This has led to the exploration of modified gravity theories as alternative frameworks capable of addressing such large-scale phenomena without invoking unknown energy components.
%

A key requirement for any cosmological model is the stability of its background evolution, typically analyzed through perturbations around a homogeneous and isotropic spacetime. These perturbations are crucial both for theoretical consistency and for explaining the formation of cosmic structures. Although general relativity (GR) succeeds on astrophysical scales, it encounters tensions on cosmological distances$\textendash\textendash$particularly in explaining the Universe's accelerated expansion. This motivates the study of modified gravity theories as alternatives that can address large-scale phenomena without invoking unknown energy components.

One promising direction in this area is massive gravity$\textendash\textendash$an effective field theory in which the spin-2 particle (graviton) propagates with a tiny but nonzero mass (see, e.g., \cite{hinterbichler2012theoretical,derham2014massive} for reviews). This theory exemplifies a typical large-scale modification of GR, which$\textendash\textendash$due to four-dimensional Poincar\'{e} symmetry$\textendash\textendash$supports five propagating degrees of freedom, in contrast to the two in GR. The most straightforward approach to introduce mass into graviton involves adding a non-derivative mass term to the Einstein-Hilbert action. At linearized level, this idea dates back to the Fierz–Pauli theory of the late 1930s \cite{fierz1939on}. More recently, the development of a fully nonlinear and Boulware-Deser ghost-free construction$\textendash\textendash$known as de Rham-Gabadadze-Tolley (dRGT) massive gravity$\textendash\textendash$was proposed in the early 2010s \cite{derham2010generalization,derham2011resummation}, providing a self-consistent and Lorentz-invariant realization of massive gravity.
Further studies on dRGT cosmology \cite{damico2011massive} uncovered significant stability issues around maximally symmetric backgrounds. These include the appearance of the Higuchi ghost at the linear level \cite{higuchi1986forbid,fasiello2012cosmological}, as well as a generic ghost instability at the nonlinear level \cite{defelice2012massive}. In response, one notable extension introduced a dynamical external scalar field to endow the graviton mass with spacetime dependence$\textendash\textendash$giving rise to the framework of mass-varying massive gravity (MVMG) \cite{damico2011massive} as illustrated in the Huang-Piao-Zhou model \cite{huang2012mass}. Unfortunately, MVMG leads to a graviton mass that becomes asymptotically negligible at late-time \cite{saridakis2013phantom}, while also exhibiting unavoidable instabilities during the early-time expansion \cite{hinterbichler2013cosmologies}.
To address these shortcomings, a promising strategy is to relax the symmetry structure of the gravitational sector by abandoning exact Lorentz invariance \cite{battye2013massive,lin2013so(3)}, as precedently discussed in ref.~\cite{rubakov2004lorentz,rubakov2008infrared}. One viable approach gives rise to a preferred time evolution that effectively reduces the full $\text{SO}(3,1)$ Poincar\'{e} group to its spatial rotation subgroup, $\text{SO}(3)$. Within this Lorentz-violating framework, several hidden features of massive gravity have been uncovered \cite{blas2009Lorentz,lin2013massive,comelli2013massive}, particularly with regard to the prospects for a consistent ultraviolet (UV) completion \cite{blas2014completing}.
In 2016, A. De Felice and S. Mukohyama proposed a new strategy for achieving healthy cosmological behavior in massive gravity, known as the \textit{minimalism program}. Its first and most important realization is the minimal theory of massive gravity (MTMG) \cite{defelice2016minimal,defelice2016phenomenology}. This theory imposes two additional nontrivial constraints into the Lorentz-violating dRGT framework, ensuring that the background cosmological equations of motion remain identical to those of dRGT, while reducing the number of propagating degrees of freedom to match that of GR. Specifically, it establishes a massive version of the Friedmann–Lema\^{i}tre equations free from the nonlinear ghost instabilities that is commonly afflict dRGT-based massive gravity. 
MTMG is also classified as a new kind of massive gravity because$\textendash\textendash$unlike standard attempts$\textendash\textendash$the nontrivial constraints modify not only potential structure but also the kinetic structure in the Lagrangian \cite{defelice2016phenomenology}.
At first glance, this development offer an intriguing cosmological prospects, as discussed in ref.~\cite{defelice2017graviton,dearaujo2021minimal,defelice2021minimal}. Unfortunately, the formulation of MTMG predicts a modified effective Newtonian gravitational constant which well-behaves only for a graviton with a negative squared mass$\textendash\textendash$though this introduces a tachyonic instability. Meanwhile, for a positive squared mass, it results in strong observational constraints \cite{defelice2022extended}.
%still suffers from unresolved instabilities, as noted in ref.~\cite{defelice2022extended}.
%
In this Letter, we show that stable cosmological solutions can be obtained by applying the minimalism program of MTMG to the framework of MVMG$\textendash\textendash$examined through the lens of cosmological perturbation theory.
%
	
%
%\section{The MTMVMG}
\textbf{The MTMVMG.} In ref.~\cite{falah2021higher} we introduced an implementation of the minimalism program into the MVMG \textit{\`{a} la} Huang-Piao-Zhou model \cite{huang2012mass}.
To set the stage for our analysis, we begin with the Arnowitt-Deser-Misner (ADM) formalism such that the physical metric and the fiducial metric can be decomposed as, $g_{\mu\nu} d x^\mu d x^\nu = - N^2 d t^2 + \gamma_{ij} (N^i d t + d x ^i) (N^j d t + d x ^j)$ and $f_{\mu\nu} d x^\mu d x^\nu = - M^2 d t^2 + \tilde{\gamma}_{ij} (M^i d t + d x ^i) (M^j d t + d x ^j)$, respectively. %, where $N$ and $M$ are the lapse functions, $N^i$ and $M^i$ are the shift vectors, and $\gamma_{ij}$ and $\tilde{\gamma}_{ij}$ are the induced spatial metrics. 
The graviton mass is generated through the condensation of an external scalar field, $\psi$, via a potential function $W = W(\psi)$ associated with the mass parameter of the graviton (see ref.~\cite{damico2011massive,huang2012mass}).
To incorporate Lorentz violation in the MVMG Lagrangian, we adopt upper-triangular vielbeins for each metric, thereby defining the so-called \textit{the precursor theory}.
We then implement the minimal theory paradigm \cite{defelice2016minimal,defelice2016phenomenology} within the Hamiltonian formulation of this precursor theory by imposing four additional constraints, $\mathcal{C}_0 \approx 0$ and $\mathcal{C}_i \approx 0$, enforced through corresponding Lagrange multipliers $\lambda$ and $\lambda^i$, respectively.
%
%modified kinetic
In particular, the constraint $\mathcal{C}_0$ contains the scalar field expression, $(\dot{\psi}-N^i\partial_i\psi) / N$, and the extrinsic curvature tensor, $K_{ij} = (\dot{\gamma}_{ij} - D_i N_j - D_j N_i) / (2 N)$, where overdot denotes the derivative with respect to time $t$, and $D_i$ is the covariant derivative with respect to the spatial metric $\gamma_{ij}$.
Thus, after integrating out $\lambda$ yields the formulation with the kinetic structure will differs from the original  precursor theory i.e. unlike the standard Einstein-Hilbert coupled to the canonical scalar field.
This also implies that, in four-dimensional spacetime, the theory propagates only three physical degrees of freedom$\textendash\textendash$consistent with Lorentz-violating massive gravity \cite{rubakov2008infrared}$\textendash\textendash$and matches the number found in scalar-tensor modified gravity \cite{ratra1987cosmological}.
This construction defines what is called the \textit{minimal theory of mass-varying massive gravity} (MTMVMG).
The effective action of MTMVMG in the unitary gauge is derived in ref.~\cite{falah2021higher}, resumed as follows
%
%\begin{widetext}
\begin{align}
	%\begin{split}
	S =&\, \int d t\, d ^3 x\, N \sqrt{\gamma}\, \Bigg\{ \frac{M_{\text{Pl}} ^{2}}{2}\, \Big( K^{i j}\, K_{i j} - K^2 + {^{(3)} R } \Big)\nonumber\\
	& + \frac{(\dot{\psi}-N^i\partial_i\psi)^2}{2 N^2} - \frac{1}{2}\, \partial^i \psi\, \partial_i \psi - V + \, W\, \bigg( \sum_{n=0}^{4} c_n \mathcal{S}_n \bigg)\nonumber\\
	& + \frac{M ^2}{N ^2} \lambda ^2\, \bigg[\bigg(\frac{W}{2M_\text{Pl}}\bigg)^2 \bigg( \Theta_{ij} \Theta^{ij} - \frac{1}{2} \Theta^2 \bigg) + \bigg(\frac{d W}{d \psi} \Phi\bigg)^2\bigg]\nonumber\\
	& - \frac{M}{N}\, \big(\lambda\, \bar{\mathcal{C}}_0 + \lambda^i\, \mathcal{C}_i \big) \Bigg\}\, ,\label{eq:mtmvmgact}
	%\end{split}
\end{align}
%\end{widetext}
%
where $\gamma = \det \gamma_{ij}$, $M_\text{Pl}$ is the reduced Planck mass, ${^{(3)}R}$ is the three-dimensional Ricci scalar associated with the spatial metric $\gamma_{ij}$, $V = V(\psi)$ denotes the self-interaction potential of the external scalar field $\psi$, and $\mathcal{S}_{n = 0, 1, 2, 3, 4}$ are potential terms inherited from dRGT massive gravity, each accompanied by a dimensionless free parameter $c_n$.
In addition, we define $\bar{\mathcal{C}}_0 \equiv \mathcal{C}_0 \big\vert_{\lambda = 0}$ and introduce functions $\Theta_{ij}$ and $\Phi$, built from $c_n$, $\gamma_{ij}$, and $\tilde{\gamma}_{ij}$.
Note that the kinetic structure deviates from the original MVMG due to the presence of the $\lambda^2$-term.
\textbf{Cosmological background and its constraint.} To satisfy the homogenous and isotropic background, we employ the Friedmann-Lema\^{i}tre-Robertson-Walker (FLRW) metric---with spatially flat---such that the physical and fiducial metrics are written by $g_{\mu\nu} d x^\mu d x^\nu = - N ^2 (t) d t ^2 + a ^2 (t) \delta _{ij} d x^i d x^j$ and $f_{\mu\nu} d x^\mu d x^\nu = - M ^2 (t)\, d t^2 + \tilde{a} ^2 (t) \delta_{ij} d x^i d x^j$, respectively, where $a (t)$ and $\tilde{a} (t)$ are their corresponding scale factors. 
The background dynamics are described in terms of the Friedmann-Lema\^{i}tre equations and the scalar field evolution. 
In this respect, we introduce the physical Hubble parameter, $H = \dot{a} /(N a)$, the fiducial Hubble parameter, $H_f = \dot{\tilde{a}} / (M \tilde{a})$, and the parameter of scale factor ratio, $u = \tilde{a} / a$.
On the other hand, we find the background constraint by evaluating the MTMVMG action \eqref{eq:mtmvmgact} with respect to $\lambda$, yields
\begin{align}
	&\frac{\lambda}{12} \frac{M}{N} \bigg(\frac{W \Theta}{M_\text{Pl}}\bigg)^2 - \lambda \frac{M}{N} \bigg( \frac{d W}{d \psi} \Phi\bigg) ^2 + \frac{( H - H_f u ) W \Theta}{2}\,\nonumber\\
	&\qquad + \frac{\dot{\psi}}{N} \frac{d W}{d \psi} \Phi = 0\, ,\label{eq:lambdaeq}
\end{align}
where we now have the formula for the functions, $\Theta = 6\, (c_1\, u^2 + 2\, c_2\, u + c_3)$ and $\Phi = c_0\, u^3 + 3\, c_1\, u^2 + 3\, c_2\, u + c_3$. 
By using the consistency conditions through the Friedman-Lema\^{i}tre equations and external scalar field evolution, a particular solution to the Lagrange multiplier has been determined as, $\lambda = 0$. 
We emphasize that the following discussions are conducted under the framework of this particular solution.
Hence, the dynamics of the cosmological background stated concisely as follows \cite{falah2021higher}
\begin{equation}
	3 M_\text{Pl} ^2 H ^2 - \rho_\text{MG} = 0\, , \qquad
	\frac{2 \dot{H}}{N} + 3H ^2 + \frac{ P_\text{MG}}{ M_\text{Pl} ^2 } = 0 \, ,\label{eq:friedmannlemaitre}%\\
\end{equation}
where
\begin{align}
	\begin{split}	
	\rho_\text{MG} &= \frac{\dot{\psi} ^2}{2N^2} + V + W (c _1 u ^3 + 3 c _2 u ^2 + 3 c _3 u + c _4 ) ,\\
	P_\text{MG} &= \frac{\dot{\psi} ^2}{ 2N ^2} - V - \frac{M}{N}\,  \frac{W \Theta}{6} - W ( c_2 u^2 + 2c_3 u + c_4 ) ,
	\end{split}
\end{align}
and
\begin{align}
	\begin{split}
	&\frac{\ddot{\psi}}{N ^2} + \bigg(3H - \frac{\dot{N}}{N^2} \bigg) \frac{\dot{\psi}}{N} + \frac{d V}{d \psi} \\
	&\quad + \frac{dW}{d\psi} \bigg(\frac{M}{N} \Phi + c _1 u ^3 + 3 c _2 u ^2 + 3 c _3 u + c _4\bigg) = 0 \, ,\label{eq:scalarconserv}
	\end{split}
\end{align}
that is perfectly identical with the MVMG cosmological background \cite{huang2012mass}.
Still, under $\lambda = 0$, there is a notable implication from the constraint eq. \eqref{eq:lambdaeq} that restrict the mass-varying potential,
\begin{equation}
	W = W_0 \exp \bigg(- \int N dt\, \frac{(H - H_f u)\, \Theta}{2\, \Phi}\bigg)\, ,\label{eq:potweq}
\end{equation}
where $W _0$ is a positive integration constant. 
The expression above bears resemblance to the mass-varying potential presented in ref.~\cite{saridakis2013phantom}. 
However, the integration in eq.~\eqref{eq:potweq} cannot be solved analytically, resulting in a distinct characteristic of our potential $W$ compared to the flat MVMG case, where $W$ asymptotically approaches zero at late-time.
Notice also that a particular feature emerges under the condition $(H - H_f u)\, \Theta = 0$ and $\Phi \ne 0$ which corresponds to either $u = H / H_f$ or $u = -\, (c_2 \pm \sqrt{c_2 ^2 - c_1 c_3}) / c_1$ leading to $W = \text{constant}$. 
We interpret this as the MTMVMG limit of the MTMG framework, where specific solutions ($H - H_f u = 0$ and $\Theta = 0$) correspond to the normal and self-accelerating branch, respectively.
%
%\section{Cosmological Perturbation}
	
\textbf{Cosmological perturbation and ghost-free conditions.} In order to address some stability issues, let us consider a set of small perturbations around the background configuration.
We define the physical metric perturbation where its perturbed components take the forms,
\begin{eqnarray}
	\begin{split}
	N &= N (t)\, \big(1 + \alpha (\textbf{x} , t) \big)\, ,\\
	N_i &= N (t)\, a (t)\, \beta_i (\textbf{x} , t)\, ,\\
	\gamma_{ij} &= a^2 (t)\, \big(\delta_{ij} + h_{ij} ( \textbf{x} , t )\big) \, .\label{eq:perturbsetup1}
	\end{split}
\end{eqnarray}
Similarly, the external scalar field and Lagrange multipliers perturbed as,
\begin{align}
	\begin{split}
	&\qquad\psi = \psi^{(0)} (t) + \delta\psi (\textbf{x} , t)\, ,\\
	&\lambda = \delta \lambda (\textbf{x} , t) \, ,\qquad \lambda _i = \delta \lambda _i (\textbf{x} , t) \, ,
	\end{split}
\end{align}
where $\psi ^{(0)} (t)$ denotes the background value of $\psi (\textbf{x},t)$.
We suppose all perturbations are in first-order, i.e. $\alpha$, $\beta_i$, $h_{ij}$, $\delta \psi$, $\delta \lambda$, $\delta \lambda_i$ $= O \,(\epsilon)$.
Then, for the sake of technical convenience we decomposed $\beta _i$, $\delta \lambda _i$, and $h _{ij}$ via the Helmholtz theorem,
\begin{align}
	\begin{split}
	&\qquad\beta_i = B_i ^\text{T} + \partial_i \beta\, ,\qquad \delta \lambda_i = \delta \lambda_i ^\text{T} + \partial_i \delta \lambda ^\text{L}\, ,\\
	&h_{ij} = A\, \delta_{ij} + \Big(\partial_i \partial_j - \frac{1}{3} \delta_{ij} \bigtriangleup\Big)\, E + 2\, \partial_i F _j ^\text{T} + h _{ij} ^\text{TT}\, ,\label{eq:helmholtzdecomp}
	\end{split}
\end{align}
where $\beta$, $\delta \lambda ^\text{L}$, $A$, $E$ are the scalar fields, $B_i ^\text{T}$, $\delta \lambda_i ^\text{T}$, $F _i ^\text{T}$ are the transverse vectors, $h _{ij} ^\text{TT}$ is the symmetric transverse-traceless tensor, $\bigtriangleup = \partial^k \partial_k$ is the Laplacian operator in the spacelike hypersurface.
Needless to say, since we work in unitary gauge, the fiducial metric $f_{\mu\nu}$ left with the unperturbed form, i.e. $f_{\mu\nu} = \text{diag} (-M^2, \tilde{a}^2, \tilde{a}^2, \tilde{a}^2)$.
After some computations, the second-order action can be decomposed into tensor, vector, and scalar modes. 
Firstly, we begin our analysis with the tensor mode and present the corresponding quadratic action in momentum space as,
\begin{align} 
	S^{(2)}_\text{ten} = \frac{M_\text{Pl}^{2}}{8} \int d t\, d^3 k\, N a^3 \bigg(\frac{\vert\dot{h} _{ij} ^\text{TT} \vert^2}{N ^2} - \omega_\text{ten} ^2 \vert h _{ij} ^\text{TT} \vert^2 \bigg) \, ,\label{eq:acttensormtmvmtg}
\end{align}
where $\omega_\text{ten}^2 = (k^2/a ^2) + M _\text{GW} ^2$ is the dispersion relation of the tensor mode, and
\begin{equation}
	M_\text{GW} ^2 = \frac{W^{(0)} u }{M_\text{Pl} ^2} \bigg[ c_2\, u + c_3 + \frac{M}{N}\, (c_1 u + c_2 ) \bigg] ,
\end{equation}
is the mass parameter of the massive gravitational waves with $W ^{(0)} = W (\psi ^{(0)})$. %$\text{r} = M / (N u)$
This result is in agreement with the standard massive gravity perturbation (see e.g. \cite{gumrukcuoglu2011cosmo}), which reveals the time-dependent mass on the propagation of the tensor mode.
To avoid the tachyonic instability, it is necessary for the mass parameter to be positive definite, $M_\text{GW} ^2 \ge 0$.
In addition, it is also safe to restrict the magnitude of $M_\text{GW}$ is sufficiently small at late-time$\textendash\textendash$at least comparable with the physical Hubble parameter, $H$.
This bound is important due to some constraints from the gravitational waves observations \cite{Ligoscientific2020test,derham2017grav}.
Secondly, we examine the vector modes, which are composed of $F_i ^\text{T}$, $B_i ^\text{T}$, and $\delta \lambda_i ^\text{T}$. 
The last two modes are the auxiliary and the Lagrange multiplier, respectively. 
Besides, by evaluating the variation of $\delta \lambda_i ^\text{T}$ gives the following constraint,
\begin{equation}
	\frac{W ^{(0)}\, (c_1 u^2 + c_2 u)}{a^2}\, k^2 F_i ^\text{T}= 0\, ,
\end{equation}
where allows us to define a useful solution, $k^2 F_i ^\text{T}= 0$, by expressing $W^{(0)}$ and $c_1 u^2 + c_2 u$ in their generic forms. 
As a result, all vector modes can be excluded, leading to the quadratic action of vector modes expressed as $S_\text{vec} ^{(2)} = 0$.
Lastly, we consider the second-order action of scalar mode which is comprised of $\alpha$, $\delta \psi$, $\delta \lambda$, $\beta$, $\delta \lambda ^\text{L}$, $A$, and $E$. 
However, only $\delta \psi$, $A$, and $E$ exhibit dynamical properties that are present from the Einstein-Hilbert and the external scalar field action. 
The remaining modes are either auxiliaries or Lagrange multipliers, which can be integrated out by using their own equations of motion. 
In particular, we obtain the correspondence formula from the equation of motion for $\delta \lambda ^\text{L}$,
\begin{equation}
	E = \frac{3}{k^2} (A - \mathcal{J} {\delta\psi})\, ,\label{eq:aedeltapsirelation}
\end{equation}
with $\mathcal{J} = \Phi / (c_1 u^2 + c_2 u)\, (d \ln W / d \psi) \vert_{\psi = \psi ^{(0)}}$.
%
	%\begin{equation*}
	%\mathcal{J} = \frac{\Phi}{ c_1 u^2 + c_2 u}\, \frac{d \ln W}{d \psi} \bigg\vert_{\psi = \psi ^{(0)}}\, .
	%\end{equation*}
%
Using this correspondence formula, one physical mode can be reduced. 
Unfortunately, the structure of the minimalism constraints resulting the total kinetic terms of all scalar modes vanished identically. 
This result leads to a \textit{nondynamical problem}, a feature commonly encountered in various massive gravity models \cite{gumrukcuoglu2011cosmo,comelli2012degrees,comelli2013massive,comelli2014nonderivative}.
Accordingly, in order to avoid the non-dynamical problem, we propose an \textit{ans\"{a}tze} by giving the specific form to the parameter of scale factor ratio,
{\allowdisplaybreaks
\begin{align}
	u = \frac{\tilde{a}}{a} = \bigg(1 - \frac{\dot{\psi} ^{(0) 2}}{6 M_\text{Pl} ^2 N ^2 H ^2}\bigg) \frac{H}{H_f}\, .\label{eq:psidotconstr}
\end{align}
This configuration$\textendash\textendash$without spoiling background equations of motion$\textendash\textendash$technically protects the kinetic terms of $\delta \psi$, $A$, and $E$ from self-cancellation due to the perturbation of minimalism constraints.}
At the same time, this is also physically significant to the physical and the fiducial metrics where they are glued to each other via their time derivative of the scale factors.
Evaluate the equation of motion for $\delta \lambda$, we find
%
%\begin{equation}
%	A = \mathcal{F} (k, t)\, \delta \psi\, ,\label{eq:adeltapsirelatin}
%\end{equation}
\begin{widetext}
\begin{align}
	&\Bigg[\dfrac{\dot{\psi} ^{(0) 3}}{2 N^3 H^3} \dfrac{\Theta}{M_\text{Pl}^2} + \dfrac{\dot{\psi} ^{(0) 2} (\Theta + 6 \Phi)}{N^2 H^2} \dfrac{d \ln W}{d \psi} \bigg\vert_{\psi = \psi^{(0)}} - \frac{\Theta}{H^2} \frac{d V}{d \psi} \bigg\vert_{\psi\, =\, \psi^{(0)}} - \frac{\Theta\, \Upsilon}{H^2} \frac{d W}{d \psi} \bigg\vert_{\psi\, =\, \psi^{(0)}} - \frac{12 M_\text{Pl}^2 \dot{\psi}^{(0)}}{N H} \frac{\Phi}{W^{(0)}} \frac{d^2 W}{d \psi^2} \bigg\vert_{\psi\, =\, \psi^{(0)}}\nonumber\\
	&\qquad + \dfrac{M_\text{Pl}^2 \mathcal{J} \Theta k^2}{2 H^2 a^2}\Bigg] \delta\psi + \Bigg[ \dfrac{3 \dot{\psi} ^{(0) 2}}{N^2 H^2}\, (2 c_2 u + c_3) + \frac{3 W^{(0)} \Theta (c_2 u^2 + c_3 u + c_4)}{H^2} + \dfrac{18 M_\text{Pl}^2 \dot{\psi} ^{(0)} (c_1 u^2 + 2c_2 u)}{N H} \dfrac{d \ln W}{d \psi} \bigg\vert_{\psi = \psi^{(0)}}\nonumber\\
	&\qquad - 9 M_\text{Pl}^2 \Theta\Bigg] A = 0\, .\label{eq:adeltapsirelatin}
\end{align}
\end{widetext}
where $\Upsilon = c_1 u^3 + 3 c_2 u^2 + 3 c_1 u + c_4$.
By using eq.~\eqref{eq:aedeltapsirelation} and eq.~\eqref{eq:adeltapsirelatin}, therefore we are left with only one active mode in the scalar perturbation, and its quadratic action in the momentum space can be expressed as,
\begin{equation}
	S^{(2)} _\text{sc} = \frac{1}{2} \int d t\, d^3 k\, N a ^3\, \bigg(\mathcal{G} ^2\, \frac{\delta \dot{\psi} ^2}{N ^2} - \omega_\text{sc} ^2 \, \delta \psi  ^2 \bigg)\, ,\label{eq:secondscalarmode}
\end{equation}
where
\begin{equation}
	\mathcal{G}^2 = 1 + \frac{\dot{\psi} ^{(0) 2}\, \Theta}{24 M_\text{Pl} ^2 N ^2 H^2\, (c_1 u^2 + c_2 u)}\, \label{eq:gfunction}
\end{equation}
is the kinetic coefficient and $\omega_\text{sc} ^2 = (c _s ^2 k ^2/a ^2) + M_\text{sc} ^2$ is the dispersion relation of the scalar mode. 
Here, $c_s$ denotes the effective sound speed of the cosmological fluid, while $M_\text{sc}$ represents the mass parameter of the scalar mode (detail expressions for these quantities are provided in eq.~\eqref{eq:soundspeedparameter} and eq.~\eqref{eq:massscalarmodeparameter} in the Appendix). 
We obtain the kinetic coefficient $\mathcal{G} ^2$ clearly always positive, and stability conditions perfectly satisfied. 
Moreover, we can avoid the divergent and tachyonic instabilities by imposing $c _s ^2 \ge 0$ and $M_\text{sc} ^2 \ge 0$, respectively.
\textbf{Phenomenology.} With the stable cosmological equations at hand, it is important to highlight some brief phenomenological applications of MTMVMG.
Following the argument in ref.~\cite{huang2012mass}, the additional scalar field $\psi(x)$ can be interpreted as quintessential dark energy in the late Universe or as the inflaton field in the early Universe.
Remember that the original motivation for considering massive gravity theories lies in their potential to provide an alternative explanation for late-time cosmic acceleration.
To address this issue, we examine the following cosmological parameters,
	\begin{equation}
		%\begin{split}
		%w _\text{MG} = \frac{P _\text{MG}}{\rho_\text{MG}} = \frac{\dot{\psi}^2 - 2 V - 2 W (c _1 u ^3 + 3 c _2 u ^2 + 3 c _3 u + c _4)}{\dot{\psi}^2 + 2 V + 2 W (c _1 u ^3 + 3 c _2 u ^2 + 3 c _3 u + c _4)} .
		\Omega_\text{DE} = \frac{\rho_\text{MG}}{3 M_\text{Pl}^2 H^2}\, , \qquad w_\text{DE} = \frac{P_\text{MG}}{\rho_\text{MG}} ,\label{eq:cosmologicalparamter}
		%\end{split} \, , \quad q = - 1 - \frac{\dot{H}}{N H^2}\,
	\end{equation}
which correspond to the dimensionless density parameter and the equation-of-state parameter, respectively, associated with the dark energy sector. 
More specifically, $\Omega_\text{DE}$ and $w_\text{DE}$ represent the total contribution arising from the massive graviton terms in MTMVMG, which are interpreted as a possible alternative to conventional dark energy.
Additionally, we will evaluate the deceleration parameter, given by $q = (1 + 3 w_\text{DE}\, \Omega_\text{DE})/2$.
Let us focus on the unperturbed background such that, $\psi = \psi^{(0)} (t)$, and$\textendash\textendash$along with this$\textendash\textendash$take the mass parameter and the self-interaction potential to be exponential forms, $W (\psi) = W_0 \exp(- \lambda_W \psi/ M_\text{Pl}^2)$ and $V (\psi) = V_0 \exp(- \lambda_V \psi/ M_\text{Pl}^2)$, respectively, where $W_0$, $V_0$, $\lambda_W$ and $\lambda_V$ are non-negative constant.
%
%We also assume that the ratio of the lapse functions $N/M$ is proportional to $u$, such that there is a dimensionless constant $\mathsf{k}$ relates them, $N/M = \mathsf{k} u$.
In the meantime, since one can choose the fiducial parameters, here, for the sake of later purpose, we consider a specific time-dependent relation between the fiducial and physical functions given by a relation $\tilde{a}/M = \mathsf{k}\, a/N$, leading to $M/N = u/\mathsf{k}$ condition, where $\mathsf{k} = \mathsf{k}(t)$ is a positive definite and dimensionless parameter that can be determined phenomenologically from the observational data.
Moreover, from the constraint in eq.~\eqref{eq:potweq} and the \textit{ans\"{a}tze} in eq.~\eqref{eq:psidotconstr}, we obtain the relation $\lambda_W \dot{\psi}/N - (\Theta / 12 H \Phi) (\dot{\psi}^2/N^2) = 0$, which yields two branch solutions.
For convenience, we refer to these as the \textit{static branch}, $\dot{\psi}/N = 0$, and the \textit{dynamic branch}, $\dot{\psi}/N = 12 \lambda_W H \Phi / \Theta$.
The static branch corresponds to the normal branch of MTMG \cite{defelice2016phenomenology}, characterized by a constant graviton mass and the condition $H_f u = H$.
In contrast, the dynamic branch allows for less constrained graviton mass, as $\psi$ remains dynamical.
%

%
%\begin{widetext}
\begin{figure*}[t]
	%\begin{subfigure*}{6cm}
	\includegraphics[width=8.9cm]{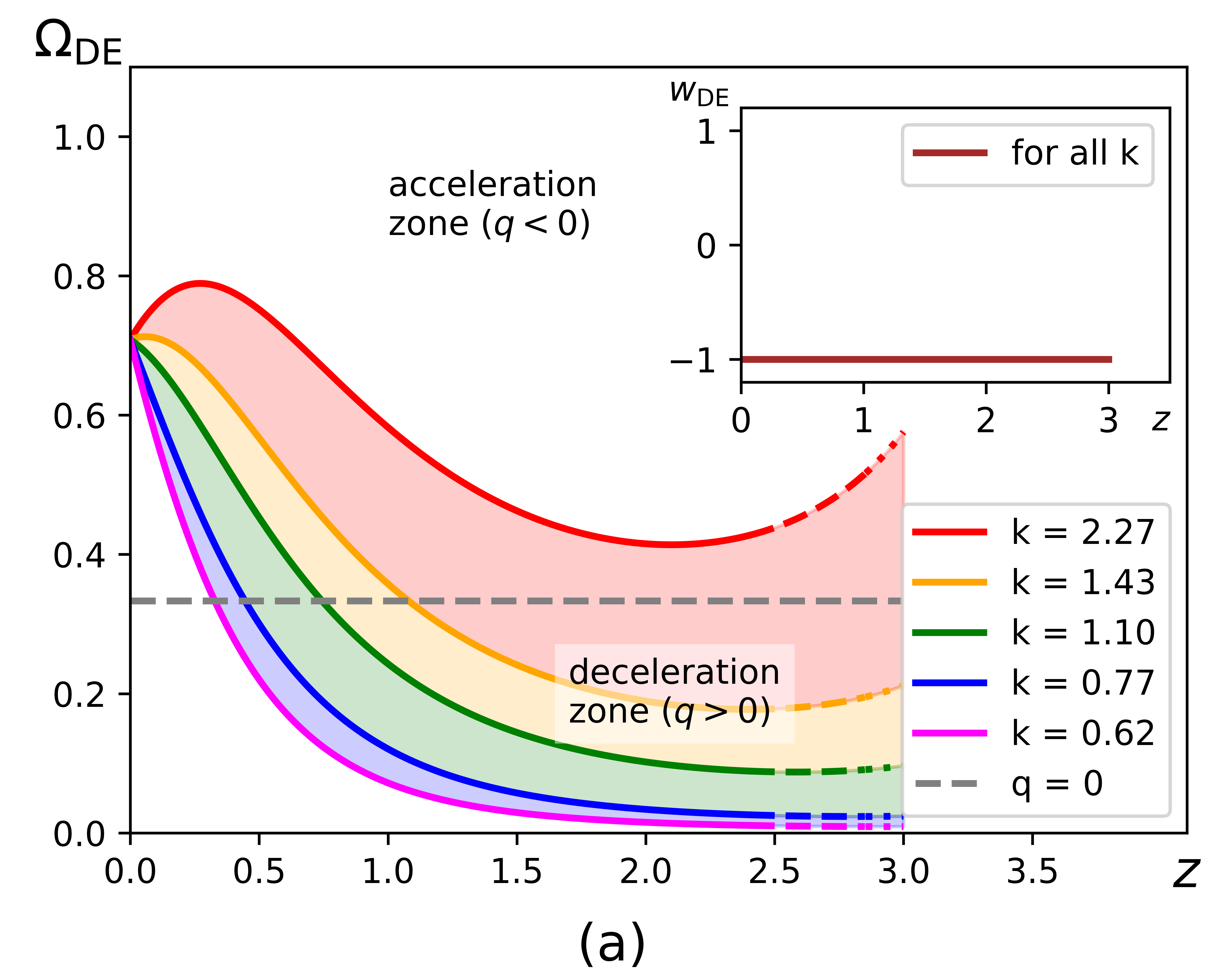}
	%\end{subfigure*}
	%
	%\begin{subfigure*}{6cm}
	\includegraphics[width=8.9cm]{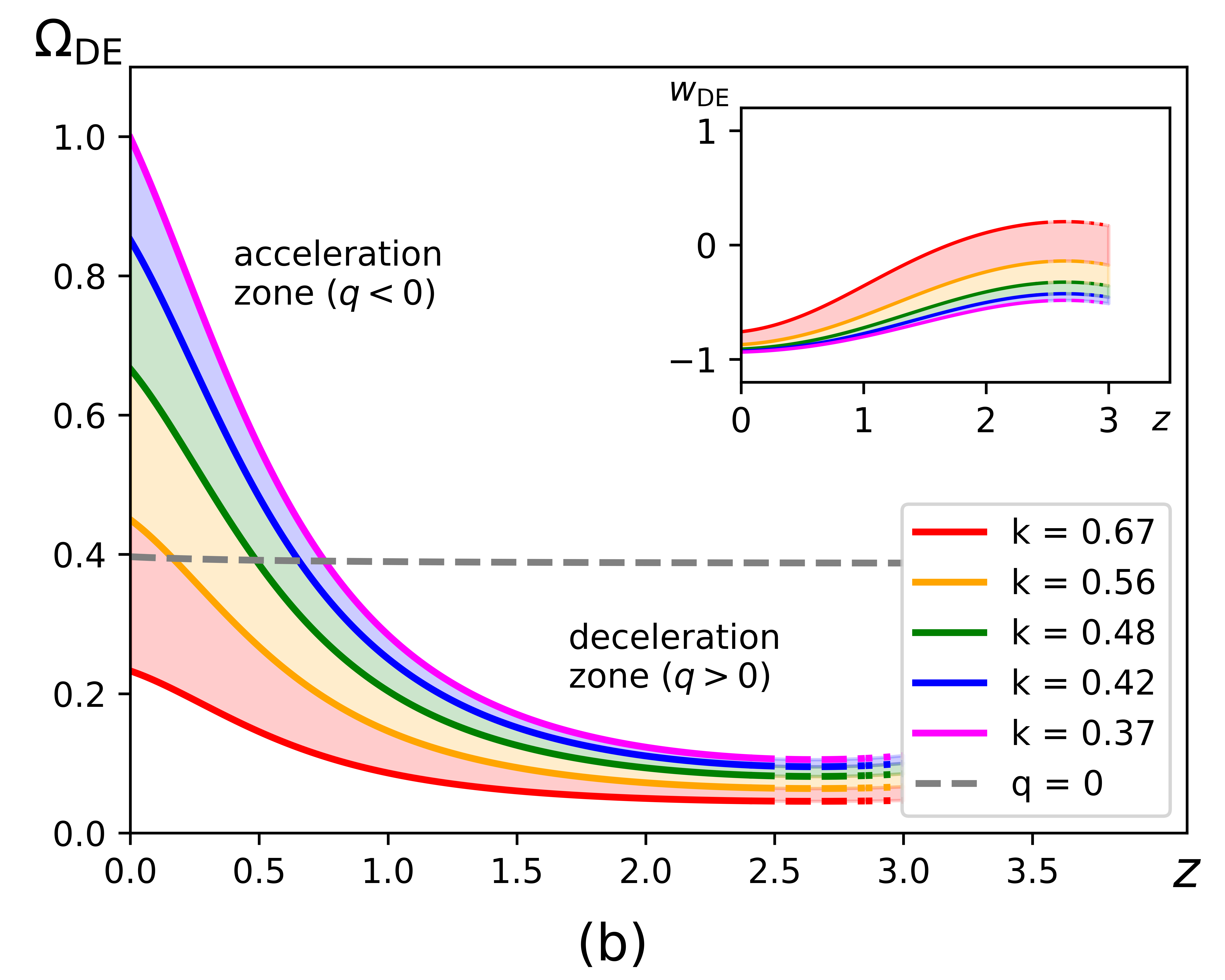}
	%\end{subfigure*}
	%\includegraphics[width=8cm]{eqofstatelatetimenew6.png}
	%
	%\includegraphics[width=8cm]{deceleratelatetimenew6.png}
	%
	\caption{Evolutions of dimensionless density parameter of dark energy $\Omega_\text{DE}$ and its corresponding equation of state parameter $w_\text{DE}$ for (a) static branch and (b) dynamic branch, shown within a certain $\mathsf{k}$ range. All dimensional parameters are normalized in unit of $M_\text{Pl} = 10$ such given that, for the static branch, $V_0 = 0.51$, $W_0 = 0.43$, $c_n = 1$, and $\lambda_{V,W}$ being arbitrary, and for the dynamic branch, $V_0 = 67.5$, $W_0 = 96.3$, $c_n = 1$, $\lambda_{V} = 0.5$, and $\lambda_{W} = 0.2$. The dashed segments represent an extrapolation beyond the fitting equation of the observed Hubble parameter, $H (z) = - 14.24 z^3 + 52.76 z^2 + 19.75 z + 68.29$, according to ref.~\cite{farooq2016hubble,salehi2022new}.}
	\label{fig:latetime}
\end{figure*}
\begin{figure}[t]
	\centering\includegraphics[width=8.5cm]{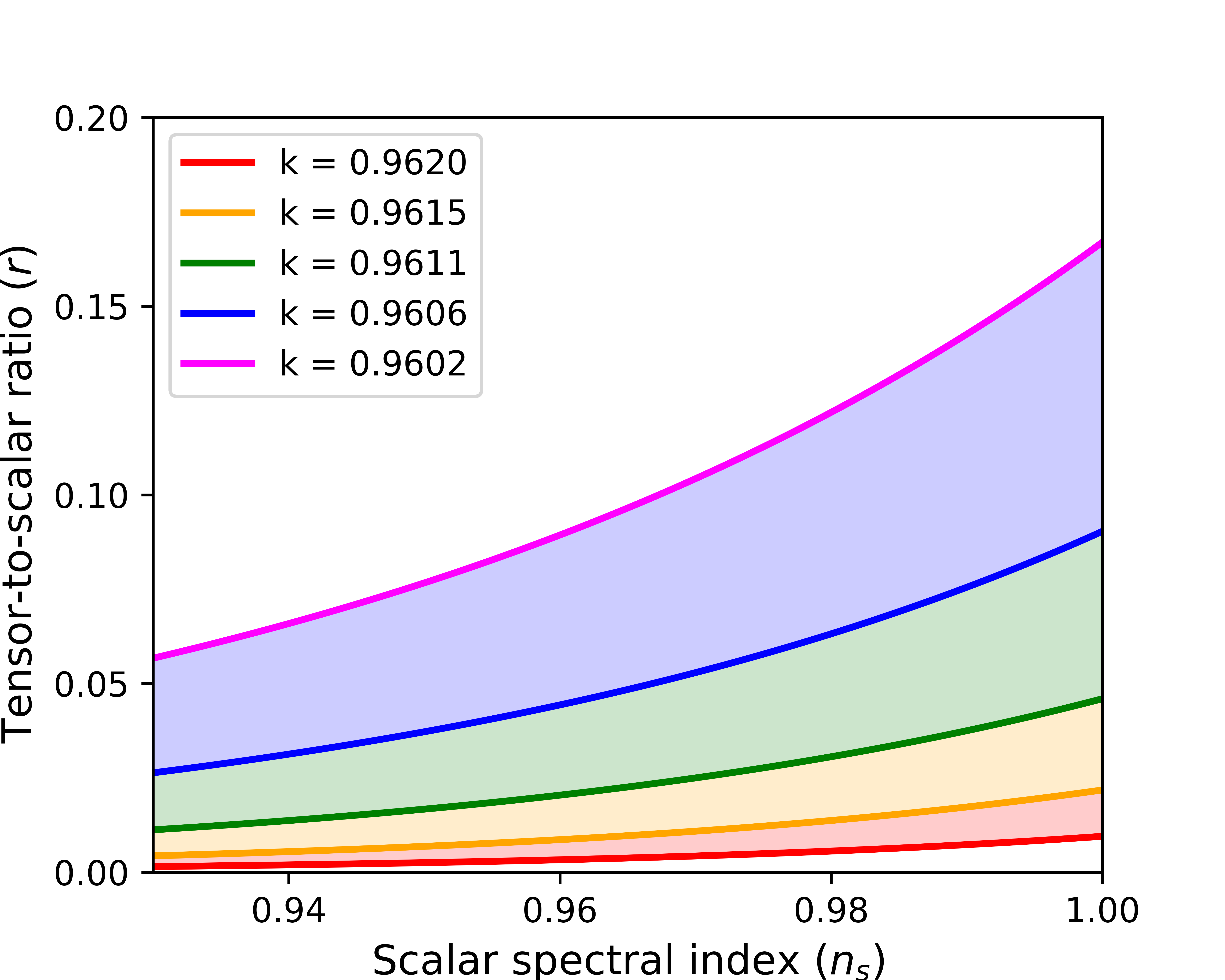}
	\caption{Plot of the tensor-to-scalar ratio ($r$) versus the spectral index of scalar mode ($n_s$) under the slow-roll approximation for the parameter range $0.9602 \le\mathsf{k}\le 0.9620$. The plot is generated by adopting exponential forms for both the mass parameter potential and the self-interaction potential, e.g. $W (\psi) = W_0 \exp (- \lambda_W \psi / M_\text{Pl})$ and $V (\psi)  = V_0 \exp (- \lambda_V \psi / M_\text{Pl})$, respectively. All dimensional parameters are normalized in unit of $M_\text{Pl} = 1$ and $H = 1$, with the coefficients adjusted to $V_0 = 0.1$, $W_0 = 0.1$, $\lambda_V = 0.1$, and $\lambda_W = 0.1$.}
	\label{fig:earlytime}
\end{figure}

%\end{widetext}
%
The evolution of $\Omega_\text{DE}$ and $w_\text{DE}$ in term of the redshift $z = a_0/a - 1$ can be demonstrated in Fig~\ref{fig:latetime}.
We emphasize that this evolution is derived under certain parameter values and using the observationally fitted Hubble parameter from ref.~\cite{farooq2016hubble,salehi2022new}, given by $H (z) = - 14.24 z^3 + 52.76 z^2 + 19.75 z + 68.29$ with a coefficient of determination $\mathfrak{R}^2 = 0.862$. 
Based on this result, we find that the static branch matches the present-day ($z = 0$) density parameter well, yielding $\Omega_{\text{DE},0} \simeq 0.7$ for $0.62 \le\mathsf{k}\le 2.27$, and exhibits a cosmological-constant-like behavior with the equation-of-state parameter, $w_{\text{DE},0} = -1$ for all $\mathsf{k}$. 
Notably, a closer inspection reveals that for $\mathsf{k} = 1.10$ the transition from deceleration to acceleration occurs at redshift, $z_\text{tr} \simeq 0.72$, in agreement with the model-independent estimate, $z_\text{tr} = 0.72 \pm 0.05$, in ref~\cite{farooq2016hubble}. 
Nonetheless, the density parameter of the dynamic branch diverges at the present day for $0.37 \le \mathsf{k} \le 0.67$, with the closest agreement to $\Omega_{\text{DE},0} \simeq 0.7$ occurring at $\mathsf{k} = 0.46$.
Although this value of $\mathsf{k}$ is appropriate, the transition from the decelerating to the accelerating phase occurs at $z_\text{tr} = 0.53$, which significantly deviates from the estimate in ref.~\cite{farooq2016hubble}.
Thus, we conclude that the static branch appears more suitable for explaining the late-time expansion than the dynamic branch.
To complete the picture of cosmic history, it is natural to explore how MTMVMG behaves in the context of the early Universe, where inflation is expected to dominate.
In this context, we describe the dynamics by treating $\psi$ as the inflaton field and introducing the slow-roll-like parameter,
\begin{align}
	\begin{split}
	\varepsilon_H &= -  \frac{\dot{H}}{N H^2} \\
	&= \frac{\dot{\psi} ^{(0)2}}{2 M _\text{Pl} ^2 N ^2 H ^2} - \frac{W ^{(0)}\, \Theta}{12 M _\text{Pl} ^2 H ^2}\, \bigg(\frac{M}{N} - u\bigg) < 1\, ,\label{eq:epsilondefined}
	\end{split}
\end{align}
thus modifies the conventional slow-roll parameter by accounting for effects arising from the graviton mass sector.
%
%This parameter is by definition similar to a standard slow-roll parameter except for one additional term raised from the graviton mass sector$\textendash\textendash$without this term of course it would return to the ordinary slow-roll parameter as in GR, $\varepsilon^\text{GR}_H = \dot{\psi} ^{(0)2} / (2 M _\text{Pl} ^2 N ^2 H ^2)$. 
%
Note that when the massless limit, $W^{(0)} = 0$, so $\varepsilon_H$ will reduces to the standard form in GR, $\varepsilon^\text{GR}_H = \dot{\psi} ^{(0)2} / (2 M _\text{Pl} ^2 N ^2 H ^2)$.
Importantly, this new parameter captures that the de Sitter geometry, characterized by $\varepsilon_H \ll 1$, can be generated without requiring the conventional slow-roll approximation, $\dot{\psi} ^{(0)2} / (2 M _\text{Pl} ^2 N ^2 H ^2) \ll 1$. 
Instead, it can be achieved by ensuring the subtraction term in the second line of eq.~\eqref{eq:epsilondefined} remains sufficiently small.
Now, let us switch from physical time, $t\, \ge 0$, to conformal time, $\eta\, \le 0$, by running the infinitesimal relation, $d \eta = N d t / a$.
At this stage, we again assume $M/N = u/\mathsf{k}$ as previously introduced in the late-time session. 
We proceed within a (nearly) de Sitter background, where the scale factor takes the form, $a (\eta) \simeq -\, 1 / H \eta$, and then impose the Bunch-Davies vacuum $\vert 0 \rangle$  in the asymptotic past ($\eta \rightarrow - \infty$).
Accordingly, the power spectra of the tensor and scalar modes can be expressed as follows,
\begin{align}
	%\begin{split}
	\mathcal{P}_h (k) &=\, \frac{2 ^{2 (1 + \mu_1)}}{\pi^3} \frac{H_\ast ^2}{M _\text{Pl} ^2}\, \vert \Gamma (\mu_{1}) \vert ^2\, \bigg(\frac{k}{k_\ast}\bigg) ^{3 - 2\, \mu_1}\, ,\\
	\mathcal{P}_{\delta \psi} (k) &=\, \frac{2 ^{2 \mu_2 - 1}}{\pi^3} \frac{c_{s \ast} ^{1 - 2 \mu_2}H_\ast ^2}{\mathcal{G} _\ast ^{3 - 2 \mu_2}}\, \vert \Gamma (\mu_{2}) \vert ^2\, \bigg(\frac{k}{k_\ast}\bigg) ^{3 - 2 \mu_2}\, ,
	%\end{split}
\end{align}
respectively.
Here, the asterisk subscript indicates the corresponding quantities are evaluated at the pivot scale, $k_\ast = 0.15\,  \text{Mpc}^{-1}$, and the gamma function $\Gamma (\mu _{1,2})$ applied for specific values of $\mu_1$ and $\mu_2$ whose their squared forms are given by,
\begin{align}
\begin{split}
	&\mu_1^2 =\, \frac{9}{4} + 2 \varepsilon_{H\ast} - \frac{M_{\text{GW} \ast} ^2}{H_\ast^2} ,\\
	\mu_2^2 =&\, \frac{9}{4} + 2 \varepsilon_{H\ast} + \frac{\dot{\mathcal{G}}_\ast}{\mathcal{G}_\ast H_\ast} - \frac{M_{\text{sc} \ast} ^2}{ \mathcal{G} _\ast ^2 H _\ast ^2} .
\end{split}
\end{align}
% 
%
%In the meantime, this Letter will put modes with a small but non-vanishing mass ($M _{\text{GW}\ast}$ $\ll H _\ast$ and $M _{\text{sc} \ast} \ll \mathcal{G} _\ast H _\ast$) lives at the slow-roll background ($\varepsilon^\text{GR}_H \ll 1$) such that $\Gamma (\mu_1) = \Gamma (\mu_2) \simeq \sqrt{\pi} / 2$.
%
From this result, one can subsequently determine the spectral indices for the tensor and scalar perturbations, respectively, which take the following form,%($c_s \approx 1$),
\begin{align}
	n_t = \frac{d \ln \mathcal{P} _h}{d \ln k} &\simeq - 2\, \varepsilon _{H\ast} + \frac{2 M _{\text{GW} \ast} ^2}{3 H _\ast ^2}\, ,\\
	n_s = 1 + \frac{d \ln \mathcal{P} _{\delta \psi}}{d \ln k} \simeq&\, 1 - \frac{4\, \varepsilon_{H\ast}}{3} - \frac{2\, \dot{\mathcal{G}}_\ast}{3\, \mathcal{G}_\ast H_\ast} + \frac{2 M _{\text{sc} \ast} ^2}{3\, \mathcal{G} _\ast ^2 H _\ast ^2}\, .
\end{align}
Note that the tensor spectral index resembles the results reported in ref.~\cite{domenech2017cmb,fujita2019blue}, which predict blue-tilted primordial gravitational waves for $M_\text{GW}^2 > 3\, \varepsilon_H H^2$$\textendash\textendash$with further discussions on the blue-tilted of tensor modes arising from massive gravity provided in ref.~\cite{wang2014inflation}.
On the other hand, the tensor-to-scalar ratio takes the form,
\begin{equation}
	r = 2^{2(\mu_1 - \mu_2) + 3}\, \frac{\mathcal{G}_\ast^{3 - 2 \mu_2}}{M_\text{Pl}^2\, c_{s\ast}^{1 - 2\mu_2}} \bigg\vert\frac{\Gamma (\mu_1)}{\Gamma (\mu_2)}\bigg\vert^2\, ,
\end{equation} 
such that we can plot $r$ against $n_s$ within the slow-roll approximation and for the range $0.9602 \le \mathsf{k} \le 0.9620$ in Fig.~\ref{fig:earlytime}.
This clearly demonstrates that MTMVMG offers a viable description consistent with the latest CMB observations, which report $n_s \simeq 0.965$ and an upper bound on the tensor-to-scalar ratio, $r < 0.056$ \cite{akrami2020planck}. 
However, behind this figure, we obtain a superluminal sound speed, $c_s^2 > 1$. 
This result can be understood as a consequence of the MTMVMG construction, which features a non-standard kinetic structure and a preferred reference frame$\textendash\textendash$both of which permit superluminal propagation, as discussed in ref.~\cite{mukhanov2006enhancing}.
Finally, several open questions remain for future investigation into the broader implications of MTMVMG. These include realizing a more complete formulation of non-slow-roll inflation, examining possible non-Gaussian features in the primordial perturbations, and developing viable screening mechanisms to satisfy local gravity tests. Further studies should also address the model predictions for primordial gravitational waves, explore its behavior under gravitational collapse, and assess its applicability to a wider range of cosmological and astrophysical phenomena.
We gratefully acknowledge  PDD  Kemendikbudristek-ITB and Riset ITB for financial support. The work of HA is partially funded by the WCR grant from Kemendikbudristek-IPB.
	
\appendix

\newpage

\begin{widetext}
\section*{Appendix}
Some detail parameters of scalar mode perturbation in action eq. \eqref{eq:secondscalarmode}
%
%
%\item Square of sound speed parameter
%
{\allowdisplaybreaks
\begin{align}
	c_s^2 =&\, 1 - \frac{\dot{\psi} ^{(0)} \mathcal{J}}{2 N H} - \frac{M_\text{Pl}^2}{4}\, \bigg(1 + \frac{\dot{H}}{NH^2}\bigg) \mathcal{J}^2  - 3 M_\text{Pl}^2 \bigg(\frac{\ddot{\psi}^{(0)}}{N^2 \mathcal{F}_2} + \frac{3\dot{\psi} ^{(0)} H}{N \mathcal{F}_2}\bigg) \Theta \mathcal{J} - \frac{3 M_\text{Pl}^2 \mathcal{J}^2 \Theta (c _2 u ^2 + c_3 u + c_4)}{4} \frac{\dot{\psi}^{(0)}}{N H \mathcal{F}_2} \frac{d W}{d\psi}\bigg\vert_{\psi = \psi^{(0)}} \nonumber\\
	& + 9 M_\text{Pl}^2 \mathcal{J}^2 (c_1 u + c_2) (c _2 u ^2 + c_3 u + c_4) \Big(1 - \frac{M}{N} \frac{H_f}{H}\Big) \frac{W^{(0)}}{\mathcal{F}_2} + \frac{3 M_\text{Pl}^2 \mathcal{J}^2 \Theta (2 c_2 u + c_3)}{4} \Big(1 - \frac{M}{N} \frac{H_f}{H}\Big) \frac{W^{(0)}}{\mathcal{F}_2}\nonumber\\
	& + \frac{3 M_\text{Pl}^2 \mathcal{J}^2 \Theta (c _2 u ^2 + c_3 u + c_4)}{4} \frac{\dot{H}}{N H^2} \frac{W^{(0)}}{\mathcal{F}_2} + \frac{3 M_\text{Pl}^2 \mathcal{J}^2 \Theta (c _2 u ^2 + c_3 u + c_4)}{4} \frac{W^{(0)} \dot{\mathcal{F}}_2}{N H \mathcal{F}_2} - \frac{3 M_\text{Pl}^2}{4} \frac{W ^{(0)}}{\mathcal{F}_2}\, (c _2 u ^2 + c_3 u + c_4) \Theta \mathcal{J}^2\nonumber\\
	& + 6 M_\text{Pl}^2\frac{\mathcal{J} \Theta}{\mathcal{F}_2} \frac{d V}{d\psi} \bigg\vert_{\psi = \psi^(0)} - 6 M_\text{Pl}^2 \bigg(c_1 \frac{M u^2}{N} + c_2 u ^2 + 2 c_2 \frac{M u}{N} - c_3 u + c_4\bigg) \frac{\mathcal{J} \Theta}{\mathcal{F}_2} \frac{d W}{d\psi} \bigg\vert_{\psi = \psi^(0)} \nonumber\\
	&- \frac{3 \dot{\psi} ^{(0)} W ^{(0)}}{N H \mathcal{F}_2} (c _2 u ^2 + c_3 u + c_4) \Theta \mathcal{J} + 3 M_\text{Pl}^2 \frac{W^{(0)} \mathcal{F}_1}{\mathcal{F}_2^2} \bigg(c_1 \frac{M u^2}{N} + c_2 u^2 + 4 c_2 \frac{M u}{N} + c_3 u + 3 c_4 \bigg) \Theta \mathcal{J} \nonumber\\
	& + \frac{ M_\text{Pl}^2}{12} \frac{W ^{(0)}}{\mathcal{F}_2} \bigg(c_1 \frac{M u^2}{N} + c_2 u^2 + c_2 \frac{M u}{N} + c_3 u\bigg) \bigg(\frac{\mathcal{F}_1}{\mathcal{F}_2} + \mathcal{J}\bigg) \Theta \mathcal{J}\, ,\label{eq:soundspeedparameter}
\end{align}}
%
%\item Square of mass parameter of scalar mode perturbation in eq. \eqref{eq:secondscalarmode}
%
{\allowdisplaybreaks
\begin{align}
	M_\text{sc}^2 =&  - \frac{1}{M _\text{Pl}^2}\, \frac{\dot{\psi}^{(0)} \ddot{\psi}^{(0)}}{N^3 H} + \frac{1}{2 M _\text{Pl}^2}\, \frac{\dot{\psi}^{(0)2} \dot{H}}{N^3 H^2} - \frac{\mathcal{J}}{2} \frac{\dot{\psi}^{(0)}}{N H} \frac{d^2 V}{d\psi^2}\bigg\vert_{\psi = \psi^{(0)}} + \frac{\mathcal{J}}{2} \frac{\dot{H}}{N H^2} \frac{d V}{d\psi}\bigg\vert_{\psi = \psi^{(0)}} + \frac{\dot{\mathcal{J}}}{2 H} \frac{d V}{d\psi}\bigg\vert_{\psi = \psi^{(0)}} - \frac{\mathcal{J} \Upsilon}{2} \frac{\dot{\psi}^{(0)}}{N H} \frac{d^2 W}{d\psi^2}\bigg\vert_{\psi = \psi^{(0)}}\nonumber\\
	& + \frac{\mathcal{J} \Upsilon}{2} \frac{\dot{H}}{N H^2} \frac{d W}{d\psi}\bigg\vert_{\psi = \psi^{(0)}} + \frac{\dot{\mathcal{J}} \Upsilon}{2 H} \frac{d W}{d\psi}\bigg\vert_{\psi = \psi^{(0)}} + \frac{\mathcal{J} \Theta}{4} \Big(1 - \frac{M}{N} \frac{H_f}{H}\Big) \frac{d W}{d\psi}\bigg\vert_{\psi = \psi^{(0)}} - \frac{\ddot{\psi}^{(0)} \mathcal{J}}{N^2} \Big(1 - \frac{\dot{\psi} ^{(0) 2}}{6 M _\text{Pl} ^2 N ^2 H ^2}\Big)\nonumber\\
	& + \Big(1 - \frac{\dot{\psi} ^{(0) 2}}{6 M _\text{Pl}^2 N ^2 H ^2}\Big) \frac{\dot{\psi} ^{(0)} \dot{\mathcal{J}}}{N^2} + \frac{\dot{\psi} ^{(0) 2} \ddot{\psi}^{(0)} \mathcal{J}}{3 M _\text{Pl}^2 N^4 H^2} - \frac{\dot{\psi}^{(0) 3} \dot{H} \mathcal{J}}{3 M_\text{Pl}^2 N^4 H^3} - \frac{3\mathcal{J}}{2} \frac{d V}{d\psi} \bigg\vert_{\psi = \psi^{(0)}} - \frac{3 \mathcal{J} \Upsilon}{2} \frac{d W}{d\psi}\bigg\vert_{\psi = \psi^{(0)}} \nonumber\\
	& - \frac{3 \dot{\psi} ^{(0)} H \mathcal{J}}{N} \Big(1 - \frac{\dot{\psi} ^{(0) 2}}{6 M_\text{Pl}^2 N^2 H^2}\Big) + \Big(\frac{3 \ddot{\psi}^{(0)}}{N^2} + \frac{9 \dot{\psi} ^{(0)} H}{N}\Big) \frac{\mathcal{F}_1}{\mathcal{F}_2} + \frac{3 \mathcal{J} (c _2 u ^2 + c_3 u + c_4)}{4} \frac{\dot{\psi}^{(0)}}{N H} \frac{\mathcal{F}_1}{\mathcal{F}_2} \frac{d W}{d\psi}\bigg\vert_{\psi = \psi^{(0)}} \nonumber\\
	& - \frac{3 \mathcal{J} (2 c_2 u + c_3)}{4} \Big(1 - \frac{M}{N} \frac{H_f}{H}\Big) \frac{W ^{(0)} \mathcal{F}_1}{\mathcal{F}_2} - \frac{3 \mathcal{J} (c _2 u ^2 + c_3 u + c_4)}{4} \frac{\dot{H}}{N H^2} \frac{W ^{(0)} \mathcal{F}_1}{\mathcal{F}_2} - \frac{3 \dot{\mathcal{J}} (c _2 u^2 + c_3 u + c_4)}{4 N H} \frac{W^{(0)} \mathcal{F}_1}{\mathcal{F}_2} \nonumber\\
	& + \frac{3 \mathcal{J} (c _2 u ^2 + c_3 u + c_4)}{4} \frac{W ^{(0)} \dot{\mathcal{F}}_1}{N H \mathcal{F}_2} - \frac{3 \mathcal{J} (c _2 u ^2 + c_3 u + c_4)}{4} \frac{W ^{(0)} \mathcal{F}_1 \dot{\mathcal{F}}_2}{N H \mathcal{F}_2^2} + \frac{3 \mathcal{J} (c _2 u ^2 + c_3 u + c_4)}{4} \frac{W ^{(0)}  \mathcal{F}_1}{\mathcal{F}_2} + \frac{d^2 V}{d\psi^2}\bigg\vert_{\psi = \psi^{(0)}} \nonumber\\ 
	& + \Big(\frac{M}{N} \Phi + \Upsilon\Big) \frac{d^2 W}{d\psi^2}\bigg\vert_{\psi = \psi^{(0)}} + \frac{\Upsilon}{M _\text{Pl} ^2} \frac{\dot{\psi} ^{(0)}}{N H} \frac{d W}{d\psi}\bigg\vert_{\psi = \psi^{(0)}} + \frac{3}{2 M _\text{Pl} ^2} \Big(1 - \frac{\dot{\psi} ^{(0) 2}}{6 M _\text{Pl} ^2 N ^2 H ^2}\Big) \frac{\dot{\psi} ^{(0) 2}}{N ^2} + \frac{\dot{\psi} ^{(0)}}{M _\text{Pl} ^2 N H} \frac{d V}{d\psi}\bigg\vert_{\psi = \psi^{(0)}} \nonumber\\
	& + \frac{6 \mathcal{F}_1}{\mathcal{F}_2} \frac{d V}{d\psi}\bigg\vert_{\psi = \psi^{(0)}} - 6 \bigg(c_1 \frac{M u^2}{N}  + c_2 u^2 + c_2 \frac{2 M u}{N} - c_3 u + c_4\bigg) \frac{\mathcal{F}_1}{\mathcal{F}_2} \frac{d W}{d\psi}\bigg\vert_{\psi = \psi^{(0)}} - \frac{3 (c _2 u ^2 + c_3 u + c_4)}{M_\text{Pl}^2} \frac{\dot{\psi} ^{(0)}}{N H} \frac{W^{(0)} \mathcal{F}_1}{\mathcal{F}_2}\, \nonumber\\
	& + 3 \bigg(c_1 \frac{M u^2}{N} + c_2 u^2 + 4 c_2\frac{M u}{N} + c_3 u + 3 c_4 \bigg)\,  \frac{W^{(0)} \mathcal{F}_1^2}{\mathcal{F}_2^2} + \frac{1}{6} W ^{(0)} \bigg(c_1 \frac{M u^2}{N} + c_2 u^2 + c_2 \frac{M u}{N} + c_3 u\bigg) \bigg(\frac{\mathcal{F}_1}{\mathcal{F}_2} + \mathcal{J}\bigg)^2 \nonumber\\
	& - \frac{\dot{\psi}^{(0) 2}}{4 N^2 H^2} \frac{\dot{\mathcal{J}}^2}{N^2} + \frac{3 \dot{\psi}^{(0)} \dot{\mathcal{J}} \mathcal{G}}{N^2} - \frac{\ddot{\psi}^{(0)}\dot{\mathcal{J}} \mathcal{G}}{2 N^3 H} - \frac{\dot{\psi}^{(0)}\ddot{\mathcal{J}} \mathcal{G}}{2 N^3 H} - \frac{\dot{\psi}^{(0)}\dot{\mathcal{J}} \dot{\mathcal{G}}}{2 N^3 H} + \frac{\dot{\psi}^{(0)} \dot{H}\dot{\mathcal{J}} \mathcal{G}}{2 N^3 H^2}\, ,\label{eq:massscalarmodeparameter}
\end{align}}
where
\begin{align}
	\mathcal{F}_1 =& \dfrac{\dot{\psi} ^{(0) 3}}{2 N^3 H^3} \dfrac{\Theta}{M_\text{Pl}^2} + \dfrac{\dot{\psi} ^{(0) 2} (\Theta + 6 \Phi)}{N^2 H^2} \dfrac{d \ln W}{d \psi} \bigg\vert_{\psi = \psi^{(0)}} - \frac{\Theta}{H^2} \frac{d V}{d \psi} \bigg\vert_{\psi\, =\, \psi^{(0)}} - \frac{\Theta\, \Upsilon}{H^2} \frac{d W}{d \psi} \bigg\vert_{\psi\, =\, \psi^{(0)}} - \frac{12 M_\text{Pl}^2 \dot{\psi}^{(0)}}{N H} \frac{\Phi}{W^{(0)}} \frac{d^2 W}{d \psi^2} \bigg\vert_{\psi\, =\, \psi^{(0)}}\, ,\\
	\mathcal{F}_2 =& \dfrac{3 \dot{\psi} ^{(0) 2}}{N^2 H^2}\, (2 c_2 u + c_3) + \frac{3 W^{(0)} \Theta (c_2 u^2 + c_3 u + c_4)}{H^2} + \dfrac{18 M_\text{Pl}^2 \dot{\psi} ^{(0)} (c_1 u^2 + 2c_2 u)}{N H} \dfrac{d \ln W}{d \psi} \bigg\vert_{\psi = \psi^{(0)}} - 9 M_\text{Pl}^2 \Theta\, .
\end{align}
		%
	%\end{itemize}
	%
\end{widetext}

\bibliography{references}

\end{document}